\documentclass[aps,prl,twocolumn]{revtex4-2}
\usepackage{amsmath,bm}
\usepackage{graphicx,epsfig,braket,amssymb,amsfonts,eufrak,tabularx,multirow,color}
\usepackage{newtxtext}
\usepackage[colorlinks=true]{hyperref}
\usepackage[T1]{fontenc}
\usepackage{textcomp}

\newcommand{\bd}{\begin{displaymath}}
\newcommand{\ed}{\end{displaymath}}
\renewcommand{\v}[1]{{\bm #1}}
\newcommand{\bpm}{\begin{pmatrix}}
\newcommand{\epm}{\end{pmatrix}}

\begin{document}

\title{Kineo-Elasticity and Nonreciprocal Phonons by Rashba-Induced Interfacial Spin-Lattice Coupling}

\author{Gyungchoon Go}
\email{gyungchoon@korea.ac.kr}
\affiliation{Division of Semiconductor Physics, Korea University, Sejong 30019, Korea}
\affiliation{Department of Physics, Korea Advanced Institute of Science and Technology, Daejeon 34141, Korea}

\author{Se Kwon Kim}
\email{sekwonkim@kaist.ac.kr}
\affiliation{Department of Physics, Korea Advanced Institute of Science and Technology, Daejeon 34141, Korea}

\begin{abstract}
We identify a previously unrecognized spin-lattice coupling allowed by broken inversion symmetry that can be viewed as a lattice analogue of electronic Rashba spin-orbit coupling. In the low-frequency regime, where magnons are integrated out, its interplay with the conventional magnetoelastic coupling generates a kineo-elastic term that couples strain and lattice velocity, giving rise to nonreciprocal phonon velocity. We further analyze the full magnon-phonon spectrum and
find that the competition between interfacial spin-lattice and magnetoelastic couplings leads to
directional hybridization and absorption, yielding asymmetric phonon propagation lengths for opposite directions. Our results indicate that such interfacial spin-lattice coupling can provide an efficient route to nonreciprocal phonon propagation in magnetic heterostructures with strong Rashba spin-orbit coupling.
\end{abstract}

\maketitle

\emph{Introduction.}\textemdash
Nonreciprocal propagation of collective excitations has attracted growing interest as a route toward directional control of energy and information flow
in condensed-matter systems~\cite{Tokura2018,Cheong2018}.
In magnetic materials, the breaking of time-reversal and spatial inversion symmetries enables nonreciprocal dynamics of spin waves and
phonons~\cite{Jamali2013, Kostylev2013, An2013, Iguchi2015, Seki2016, Sato2016, Nomura2019, Hirokane2020, Sasaki2017, XZhang2020, TYu2020, Tateno2020, Puebla2020, Rovillain2022, Cai2023}, with implications for magnonics, phononics, and spintronics.
Several microscopic origins have been reported, including long-range dipolar interaction~\cite{Jamali2013, Kostylev2013, An2013},
Dzyaloshinskii-Moriya interaction~\cite{Iguchi2015, Seki2016, Sato2016, Nomura2019}, and magnetoelastic coupling~\cite{Sasaki2017, Tateno2020, Kuss2020, Puebla2020, TYu2020, XZhang2020, Rovillain2022, Cai2023}.

Recently, it has been reported that nonreciprocal propagation can be induced by the coupling between spin and elastic rotation~\cite{Xu2020}.
This spin-vorticity coupling, expressed as $\v{m} \cdot (\nabla \times \dot{\v{u}})$, where $\v{m}$ is the magnetization and $\v{u}$ is the lattice displacement, is a universal interaction in bulk magnetic materials that exists independently of spin-orbit interaction or specific crystal symmetries~\cite{Matsuo2013, Matsuo2017, Funato2022}.
However, while this bulk term originates from the local rotation (vorticity) of the lattice motion,
the presence of a physical interface can introduce additional coupling terms.
Here, we focus on an interfacial system that emerges specifically when inversion symmetry is broken by the substrate.
This symmetry breaking defines a distinguished polar axis $\hat{\v{z}}$, allowing a local velocity-dependent spin-lattice coupling:
\begin{align}
{\cal L}_{\text{SL}} \propto \v{m} \cdot (\hat{\v{z}} \times \dot{\v{u}}).
\end{align}
which has no counterpart in centrosymmetric bulk systems.

At inversion-broken interfaces, spin-orbit coupling provides a natural microscopic origin for such a velocity-dependent spin-lattice interaction.
The Rashba spin-orbit interaction~\cite{Bychkov1984} is ubiquitous in magnetic heterostructures and has been extensively studied in the context of current-induced spin torques and chiral magnetic textures~\cite{Miron2010,Garello2013, Kurebayashi2014, KimKW2013, Yang2018, Koo2020}.
However, its coupling to lattice motion has remained largely unexplored.
This raises a fundamental question: whether and how the Rashba spin-orbit interaction can mediate spin-lattice coupling and lead to nonreciprocal phonon dynamics?

In this work, we derive such an interfacial spin-lattice coupling.
Microscopically, lattice motion modifies the chiral orbital angular momentum (OAM) texture associated with orbital Rashba physics~\cite{Petersen2000,Park2012,Park2013,PKim2014}.
Projecting the spin-orbit interaction onto the resulting lattice-modified states then yields a velocity-dependent coupling between lattice and spin degrees of freedom.
We investigate the dynamical consequences of this interfacial coupling for phonon and magnon-phonon excitations.
In the low-frequency regime, where the magnon degrees of freedom can be integrated out, the interplay between the interfacial spin-lattice coupling
and the conventional magnetoelastic coupling~\cite{Kittel1949} generates
a kineo-elastic term which couples the lattice strain with its velocity~\cite{Dong2025} in the effective phonon Lagrangian.
This term is odd under momentum reversal and leads to a nonreciprocity in transverse phonon propagation.

Beyond the low-frequency regime, we analyze the full magnon-phonon spectrum and
find that the competition between the interfacial spin-lattice coupling and the conventional magnetoelastic coupling
produces a directional hybridization between transverse phonons and magnons.
The hybridization is enhanced for one propagation direction while being suppressed for the opposite direction, leading to a marked asymmetry in phonon absorption by magnons.
Consequently, phonons are efficiently absorbed in one direction, while long-range propagation is preserved in the opposite direction.
These effects provide clear dynamical signatures of the Rashba-induced interfacial spin-lattice coupling in magnetic heterostructures with strong Rashba spin-orbit coupling.

\emph{Interfacial spin-lattice coupling.}\textemdash
To derive the interfacial spin-lattice coupling, we consider, as a
minimal microscopic realization, a $p$-orbital system at an inversion-broken interface, where hybridization between in-plane and out-of-plane orbitals generates the orbital Rashba Hamiltonian and the associated chiral OAM texture~\cite{Park2012,Park2013}. 
We decompose the lattice displacement into a center-of-mass component
$\v u_{\rm CM}$ and an internal component $\v u_i$,
with $\sum_i M_i\v u_i=0$, where $M_i$ is the mass of ion $i$.
For uniform center-of-mass motion at constant velocity, the former can be
removed by transforming to the comoving frame and does not contribute to
the internal electronic dynamics~\cite{SM}.
For a spatially smooth lattice displacement $\v u_i(t)$ varying adiabatically on the electronic timescale, the electronic states are described by the rigidly translated local basis
$|\psi_i^a(t)\rangle = e^{-i\v u_i(t)\cdot\hat{\v p}/\hbar}|p_a(\v R_i)\rangle,$
where $|p_a(\v R_i)\rangle$ with $a=x,y,z$ denotes the $p$-orbital state at site $\v R_i$.
 In this moving basis, the effective Hamiltonian acquires a time-dependent basis connection,
${\cal A}^{(t)}_{ia,jb}\approx \dot{\v u}_j\cdot \langle p_a(\v R_i)|\hat{\v p}|p_b(\v R_j)\rangle$.
While its diagonal single-band component is closely related to the conventional lattice-tracking contribution in a single-band description~\cite{Holstein1959,Khan1984,Sundaram1999},
inter-orbital matrix elements in a multi-orbital basis allow additional velocity-dependent orbital structures.

To identify the minimal local contribution, we retain the on-site
momentum matrix elements
$\langle p_a(\v R_i)|\hat{\v p}|p_b(\v R_i)\rangle$,
which are forbidden by inversion symmetry in the centrosymmetric
limit but are allowed at an inversion-broken interface.
Their orbital structure is identical to that of the inter-orbital hopping elements allowed by inversion symmetry breaking and responsible for the orbital Rashba Hamiltonian in Refs.~\cite{Petersen2000,Park2012, Park2013,PKim2014}.
Consequently, the induced velocity-dependent correction takes the form of an orbital-lattice coupling, ${\cal A}^{(t)}_{ab}\propto (\hat{\v z}\times \dot{\v u})\cdot \v L_{ab}$, where $\v L$ denotes the OAM matrices in the $p$-orbital basis, in close analogy to the orbital Rashba term.
Projecting the atomic spin-orbit interaction onto the lattice-modified eigenstates then yields, to leading order,
$H_{\rm SL}=-\Lambda_{\rm SL}\,\boldsymbol{\sigma}\cdot(\hat{\v z}\times\dot{\v u})$,
where $\Lambda_{\rm SL}\sim \hbar\alpha_R/(2ta^2)$, with $\alpha_R$ the Rashba coefficient, $t$ the ordinary hopping scale, and $a$ the lattice constant.
A microscopic derivation is given in Supplemental Material~\cite{SM}.

In the strong-exchange limit, where the electron spin dynamics is dominated by the exchange field,
the electron spin expectation value is locked to the local magnetization direction $\v m$.
Identifying $\boldsymbol{\sigma} \simeq -{\cal P} {\v m}$, where $\mathcal{P}$ is the spin polarization,
we arrive at our foundational equation for the energy density describing the effective interfacial spin-lattice coupling:
\begin{align}\label{eqSL}
{\cal H}_{\text{SL}}  = \lambda_{\text{SL}} \v m \cdot(\hat {\v z} \times \dot{\v{u}}),
\end{align}
where $\lambda_{\text{SL}} = n \mathcal{P} \Lambda_{\text{SL}}$ is the renormalized coupling constant, with $n$ the density of magnetic moments.
This velocity-dependent interaction provides a hitherto unrecognized symmetry-allowed coupling between magnetic order
and lattice dynamics and it is our first main result.

For a localized acoustic wave at the interface, characterized by a displacement field $u \propto e^{-z/l}$,
where $l$ is the penetration depth of the acoustic wave, this interaction resembles the spin-vorticity coupling~\cite{Matsuo2013,Matsuo2017,Funato2022}:
${\cal H}_{\text{sv}} = n S {\v m} \cdot (\nabla \times \dot{\v{u}})/2 = \lambda_{\text{sv}} \v m \cdot(\hat {\v z} \times \dot{\v{u}})$,
where $\lambda_{\text{sv}} = n S /(2l)$.
However, the two couplings have different microscopic origins and can differ greatly in magnitude.
Because our interaction originates from interfacial spin-orbit coupling,
unlike spin-vorticity coupling, it can be much stronger in systems with strong interfacial spin-orbit coupling.
Using the typical parameter set specified below, together with $l = 2~\mu{\rm m}$ and $S=\hbar$,
we obtain a ratio $\lambda_{\text{SL}}/\lambda_{\text{sv}} \approx 2.2\times 10^3$.
This enhancement indicates that the Rashba-induced spin-lattice coupling can be much more potent than the spin-vorticity effect at inversion-broken interfaces.
Below, we examine its effects on the nonreciprocal phonon spectrum
and magnon-phonon hybridization.

\begin{figure}[t]
\includegraphics[width=1.0\columnwidth]{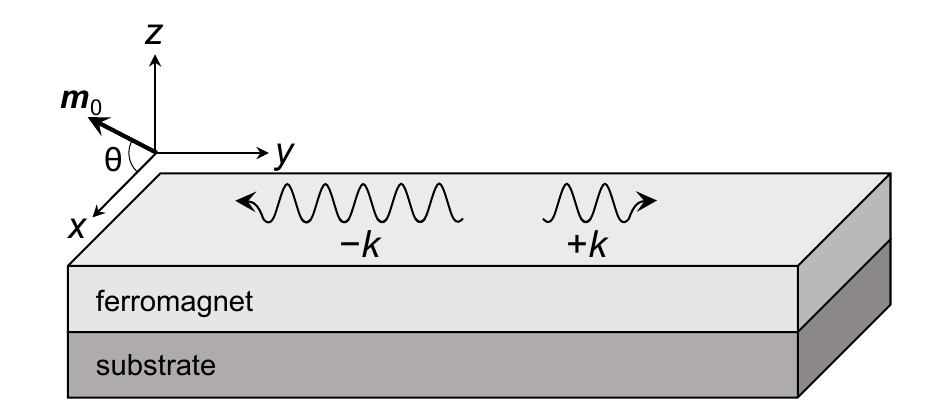}
\caption{\textbf{Model geometry.}
A two-dimensional ferromagnetic film in the $xy$-plane is bonded to a substrate with normal $\hat z$.
We focus on quasi-one-dimensional propagation along $\hat y$.
The equilibrium magnetization is ${\v m}_0 = (\cos \theta, 0, \sin \theta)$.
Wavy arrows denote acoustic waves with wavevectors $\pm k$; their unequal lengths indicate nonreciprocal phonon propagation.}\label{fig:1}
\end{figure}

\emph{Minimal model.}\textemdash
We consider a ferromagnetic thin film in the $xy$-plane, bonded to a substrate below $(z<0)$ that breaks inversion symmetry under $(z \rightarrow -z)$.
For clarity, we restrict our analysis to a quasi-one-dimensional geometry in which waves propagate along the $y$ direction
(see Supplemental Material~\cite{SM} for general in-plane propagation directions).
As shown in Fig.~\ref{fig:1}, we take the equilibrium magnetization to lie in the $xz$-plane, $\v{m}_0 = (\cos\theta, 0, \sin\theta)$, and describe small transverse fluctuations as $\delta \v{m} \approx \v{R}(\theta) (m'_x, m'_y, 0)$ with $|m'_{x,y}| \ll 1$, where $\v{R}(\theta)$ is a rotation matrix that aligns the local quantization axis $\hat {\v z}'$ with the equilibrium direction.
Hereafter, we drop the prime notation and denote the magnetization fluctuations simply by $m_x$ and $m_y$.
The internal in-plane lattice displacement is defined as $\v{u} = (u_x, u_y)$ with the corresponding strain tensor $u_{ij} = (\partial_i u_j + \partial_j u_i)/2$.

The quadratic Lagrangian density of the system is given by $\mathcal{L} = \mathcal{L}_{\text{ph}} + \mathcal{L}_{\text{m}} + \mathcal{L}_{\text{m-ph}}$:
\begin{equation}\label{L0}
\begin{aligned}
\mathcal{L} = & \underbrace{\frac{\rho}{2} \dot{\v{u}}^2 - \frac{1}{2} \v{u} \cdot \v D \cdot \v{u} }_{\mathcal{L}_{\text{ph}}} \\
& + \underbrace{\frac{\rho_s}{2} \epsilon_{ij} m_i \dot{m}_j - \frac{1}{2}\left[\Delta (m_i)^2 + A_{\text{ex}} (\nabla m_i)^2\right]}_{\mathcal{L}_{\text{m}}} \\
& - \underbrace{\lambda_\text{SL} (m_y \dot{u}_x - \sin\theta m_x \dot{u}_y)}_{\mathcal{L}_{\text{SL}}} -\underbrace{2 b_2 \cos\theta \, m_y u_{xy}}_{\mathcal{L}_{\text{me}}}
\end{aligned}
\end{equation}
where the Einstein summation convention is applied to repeated indices $i, j \in \{x, y\}$.
Here, $\rho$ is the mass density, ${\v D}_{ij} \equiv (\lambda_L + \mu) \partial_i \partial_j + \mu \nabla^2 \delta_{ij}$ is the isotropic elastic operator with Lam\'{e} coefficients $(\lambda_L, \mu)$.
In the magnon sector $\mathcal{L}_{\text{m}}$, $\rho_s$ is the spin density, $\epsilon_{ij}$ is the 2D Levi-Civita symbol dictating gyroscopic precession,
$\Delta (= \gamma \rho_s B_{\text{eff}})$ is the magnonic gap, and $A_{\text{ex}}$ is the exchange stiffness.

The interaction term $\mathcal{L}_{\text{m-ph}}=\mathcal{L}_{\text{SL}}+\mathcal{L}_{\text{me}}$ includes the interfacial spin-lattice coupling and the conventional magnetoelastic interaction~\cite{Kittel1949}. In the quasi-one-dimensional geometry considered here, we retain only the shear-strain term $u_{xy}$ with coefficient $b_2$
\footnote{The magnetoelastic coupling also contains a diagonal-strain term,
$\mathcal{L}_\text{me}^\text{diag} \approx -b_1\!\left(\cos^2\theta + m'_x \sin 2\theta \right)u_{xx}$.
The first term describes a strain-induced local canting of the magnetization direction, with $\delta\theta\sim b_1u_{xx}/\Delta$,
which is expected to be small in the regime considered here and is therefore neglected within the present quadratic fluctuation theory.
The second term gives a bilinear coupling between $u_{xx}$ and the magnon fluctuation $m_x$, but it does not contribute to transport along the $y$-direction
considered in the quasi-one-dimensional setup here.}.
For clarity, we omit both the long-range dipolar interactions and the Dzyaloshinskii-Moriya interaction (DMI) from the minimal model in Eq.~\eqref{L0}.
In the thin-film geometry considered here, dipolar-induced nonreciprocal excitations are expected to be subdominant.
The role of DMI is discussed in the Supplemental Material~\cite{SM}.

For numerical analysis, we use material parameters representative of interfacial ferromagnets such as Pt/Co:
$\alpha_R = 1~{\text{eV}\,\rm\mathring{A}}$, ${\cal P}=0.5$~\cite{Miron2010}, $t=0.5$ eV, and $a=3~\rm\mathring{A}$.
For the elastic sector, we use $\rho = 8.8$ ${\text{g/cm}}^3$, $\mu = 82$ GPa, and $\lambda_L = 135$ GPa~\cite{Mcskimin1955}.
The magnetic parameters are $\rho_s = 7.2\times 10^{28}$~$\hbar/{\text{m}}^3$ and $A_{\text{ex}} = 2.1 \times 10^{-11}$~J/m~\cite{Eyrich2012}, and the magnetoelastic coupling is $b_2 = -1.62\times 10^8$~erg/cm$^3$~\cite{Loser2000}.

\emph{Effective phonon model: kineo-elastic coupling.}\textemdash
In the adiabatic regime $\rho_s |\omega|, A_\text{ex} {k}^2 \ll \Delta$, the massive magnon fields $m_i$ can be integrated out to quadratic order.
To leading order in $\Delta^{-1}$, we derive the effective phonon Lagrangian
$\cal L_{\text{eff}} = {\cal L}_{\text{ph}} + \delta {\cal L}$, where the correction term $\delta \mathcal{L}$ is given by:
\begin{equation}\label{eqke}
\begin{aligned}
\delta {\cal L} &\simeq \frac{(\lambda_\text{SL} {\dot u}_x + 2 b_2 \cos\theta u_{xy} )^2 + (\lambda_\text{SL} \sin\theta {\dot u}_y)^2}{2\Delta}\\
&=\underbrace{\eta {\dot u}_x u_{xy}}_{\mathcal{L}_{\text{ke}}} \!+ \frac{\lambda_\text{SL}^2}{2\Delta}({\dot u}_x^2 \!+\!\sin^2\theta {\dot u}_y^2)
\!+\! \frac{2b^2_2 \cos^2\theta}{\Delta} u_{xy}^2,
\end{aligned}
\end{equation}
with the coefficient $\eta \equiv 2\lambda_\text{SL} b_2\cos\theta/\Delta$.
The last two terms in Eq.~\eqref{eqke} simply renormalize the inertia
of $u_{x,y}$ and the shear modulus; however, the term ${\mathcal{L}_{\text{ke}}}$ introduces a distinct mixed structure.
This term corresponds to the \textit{kineo-elastic coupling}, which couples the lattice strain to its velocity~\cite{Dong2025}.
In momentum space, it is odd in both the wavevector and the lattice velocity.
The distinctive feature of the present mechanism is that
it is microscopically generated by the interplay between the interfacial spin-lattice coupling and the magnetoelastic coupling~\cite{Kittel1949},
and is directly tunable by the magnetization orientation.

\begin{figure}[t]
\includegraphics[width=1.0\columnwidth]{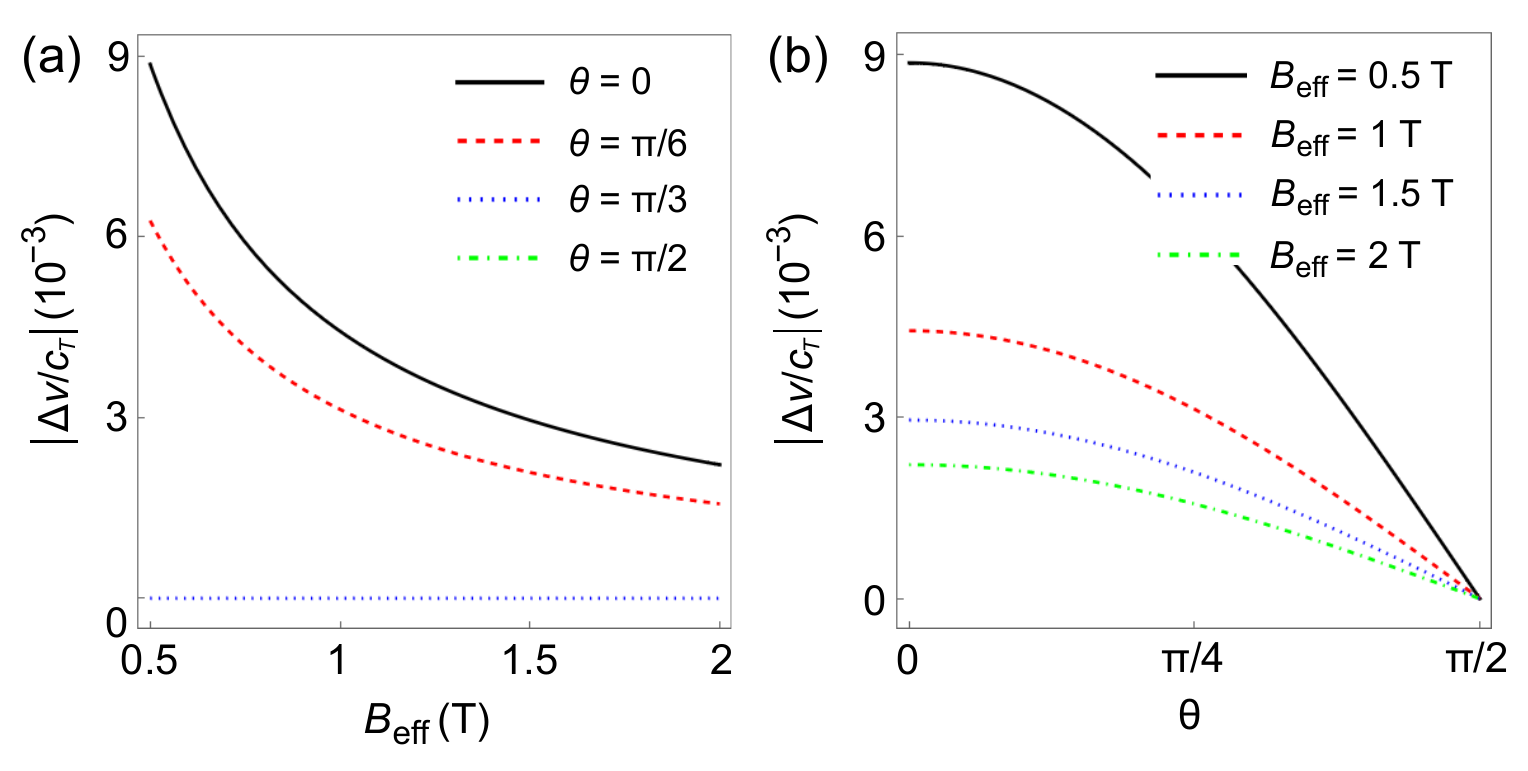}
\caption{\textbf{Nonreciprocity of the phonon velocity.}
The relative velocity asymmetry $|\Delta v / c_T|$  for waves propagating along the $y$ direction is shown as a function of (a) the effective magnetic field $B_{\text{eff}}$ and (b) the magnetization orientation $\theta$. }\label{fig:2}
\end{figure}

We now analyze the phonon modes of $\mathcal{L}_{\text{ph}} + \mathcal{L}_{\text{ke}}$, neglecting the inertia and shear-modulus renormalizations proportional to $\lambda_{\text{SL}}^2$ and $b_2^2$. For plane waves $\v {u}(\v{r},t) = \v{u} e^{i(\v{k}\cdot \v{r} - \omega t)}$, the equations of motion take the matrix form:
\begin{equation}
    [\rho\omega^{2} {\v 1} - \omega \v{K}(\v k) - {\v D}(\v k)] {\v u} = 0, \label{eq:eom_matrix}
\end{equation}
where ${D}_{ij}(\v k) = (\lambda_L + \mu)k_i k_j + \mu k^2 \delta_{ij}$,
and the kineo-elastic contribution $\v K(\v k)$ is given by:
\begin{equation}
    \v K(\v k) = \eta \begin{pmatrix} k_y & k_x/2 \\ k_x/2 & 0 \end{pmatrix}. \label{eq:K_matrix}
\end{equation}
Because $\v K(\v k)$ is odd in $\v k$, the phonon dispersions obtained from Eq.~\eqref{eq:eom_matrix}
generally satisfy $\omega_\alpha(\v k) \neq \omega_\alpha(-\v k)$.
This nonreciprocity of the phonon modes is most transparent for propagation along the $\hat{\v y}$ direction $(k_x=0)$, corresponding to the quasi-one-dimensional geometry considered here, where the longitudinal ($u_y$) and transverse ($u_x$) motions decouple [see Supplemental Material~\cite{SM}].
Keeping the leading order in $\eta$, the dispersions of the transverse and longitudinal modes are given by:
\begin{equation}
    \omega_T(k) \simeq c_T|k| + \frac{\eta}{2\rho}k, \quad \omega_L(k) = c_L|k|,\label{eq:omega_T}
\end{equation}
where $c_T = \sqrt{{\mu}/{\rho}}$ and $c_L = \sqrt{{(\lambda_{L} + 2\mu)}/{\rho}}$.
Equation \eqref{eq:omega_T} is explicitly nonreciprocal for transverse mode $\omega_T(k) \neq \omega_T(-k)$ while the longitudinal mode is reciprocal $\omega_L(k) = \omega_L(-k)$.
Consequently, the group velocity of the transverse phonon mode differs for right- and left-moving waves by $\Delta v \equiv v_g(k>0) - v_g(k<0) = \eta/\rho$.
This is our second main result: the interfacial spin-lattice coupling [Eq.~\eqref{eqSL}] induces the kineo-elastic coupling [Eq.~\eqref{eqke}]
between the lattice strain and the lattice velocity and thereby gives rise to the nonreciprocity in the transverse phonon transport~[Eq.~\eqref{eq:omega_T}; Fig.~\ref{fig:2}].

Figure~\ref{fig:2} shows the nonreciprocity of the phonon velocity.
It is maximal for $\theta = 0$ ($\v m = \hat{\v x}$),
decreases at intermediate angles, and vanishes for magnetization perpendicular to the propagation direction ($\theta = \pi/2$; $\v m = \hat{\v z}$).
This behavior is similar to nonreciprocal propagation from magnetoelastic interactions reported previously~\cite{Sasaki2017, Rovillain2022}.
A key distinction, however, is that our effective phonon model applies far from the magnon-phonon resonance,
where the magnon degrees of freedom can be integrated out.
Even off resonance, the resulting nonreciprocity can exceed the velocity asymmetry
$\Delta v/v_0 \sim 10^{-5}$ expected from magnetoelastic interactions~\cite{Sasaki2017, Rovillain2022},
depending on the Rashba strength.
The full magnon-phonon band structure, including the resonance region, is discussed below.

\emph{Directional magnon-phonon coupling.}\textemdash
We now analyze the coupled magnon and phonon modes of Eq.~\eqref{L0}.
The equations of motion can be written as
\begin{equation}\label{eqMP}
\begin{aligned}
    [\rho\omega^{2} \mathbb{P}_{\text{ph}} - i \omega \rho_s \mathbb{P}_{\text{m}} \epsilon - {\mathbb{D}}(\v k) + {\mathbb{K}}(\v k, \omega)]\Psi = 0,
\end{aligned}
\end{equation}
where $\Psi = (u_x, u_y, m_x, m_y)$.
Here $\mathbb{P}_{\text{ph}} = {\rm diag}(1,1,0,0)$ and
$\mathbb{P}_{\text{m}} = {\rm diag}(0,0,1,1)$ project onto the phonon and magnon subspaces, respectively, and $\epsilon$ is the 2D Levi-Civita tensor acting in the magnon space.
The dynamical matrix ${\mathbb{D}}(\v k)$ describes the bare phonon and magnon sectors, while
the antisymmetric matrix ${\mathbb{K}}(\v k, \omega)$ contains the magnetoelastic and interfacial spin-lattice couplings
[see Supplemental Material~\cite{SM} for full expressions].

\begin{figure}[t]
\includegraphics[width=1.0\columnwidth]{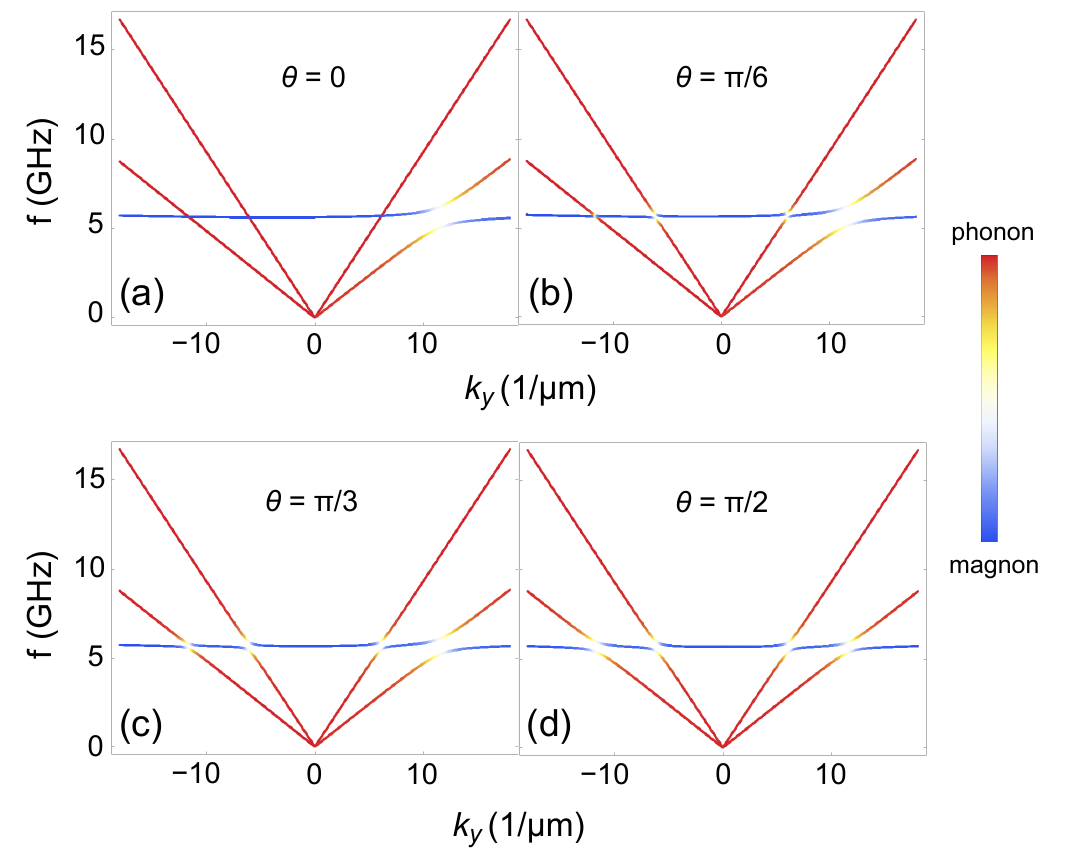}
\caption{\textbf{Directional magnon-phonon hybridization.}
Magnon-phonon dispersions versus $k_y$ for magnetization angles
(a) $\theta=0$, (b) $\theta=\pi/6$, (c) $\theta=\pi/3$, and (d) $\theta=\pi/2$.
The color scale shows the phonon (red) and magnon (blue) character of each mode.
We use $B_{\text{eff}}=0.2$~T.
}\label{fig:3}
\end{figure}

\begin{figure}[t]
\includegraphics[width=1.0\columnwidth]{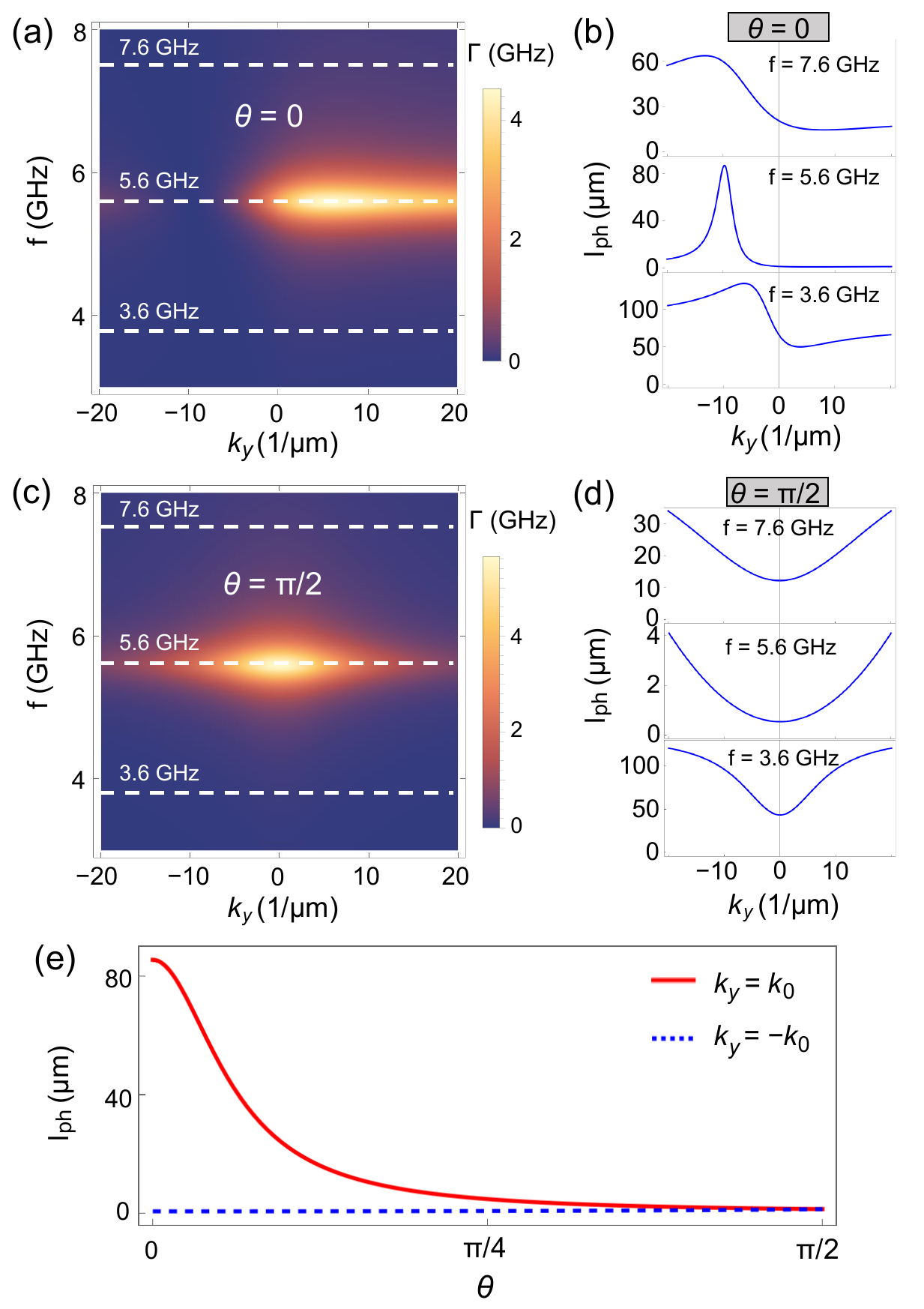}
\caption{\textbf{Nonreciprocal phonon absorption and propagation length.}
Magnon-induced phonon absorption rate $\Gamma_{\text{m-ph}}$ as a function of $k_y$ for different phonon frequencies at (a) $\theta=0$ and (c) $\theta=\pi/2$.
Panels (b) and (d) show the corresponding propagation lengths.
Panel (e) shows the magnetization-orientation dependence of the propagation length at $k_y=\pm k_0$ and $f=5.6$ GHz, with $k_0=\lambda_\text{SL}\omega/b_2$.
An effective magnetic field $B_{\text{eff}} = 0.2$~T, Gilbert damping $\alpha = 0.05$, and $Q = 1000$ are used.
The data are obtained for $k_x = 0$ and $u_x = u_y$.
}\label{fig:4}
\end{figure}

Figure~\ref{fig:3} shows the magnon-phonon spectrum along $\v k = k_y\hat{\v y}$ for different magnetization orientations.
For small $\theta$, the transverse phonon and magnon exhibit pronounced directional hybridization, with an anticrossing gap asymmetric under $k_y \rightarrow -k_y$.
This asymmetry is gradually suppressed as the magnetization is rotated and vanishes at $\theta = \pi/2$.
In contrast, no such asymmetric hybridization is observed between the longitudinal phonon and the magnon.
This behavior can be traced back to the structure of the matrix ${\mathbb{K}}$, whose only
nonvanishing elements are $\mathbb{K}_{(u_x,m_y)}$ and $\mathbb{K}_{(u_y,m_x)}$ [see Supplemental Material~\cite{SM}].
Because the hybridization between transverse phonon ($u_x$) and the magnon is governed by $\mathbb{K}_{(u_x,m_y)} = i \left(\omega \lambda_{\text{SL}} - b_2 \cos\theta k_y\right)$,
the magnetoelastic and interfacial spin-lattice couplings compete as a functions of $k_y$
resulting in directional hybridization.
By contrast, the hybridization between longitudinal phonon ($u_y$) and the magnon, $\mathbb{K}_{(u_y,m_x)} = -i \omega \lambda_{\text{SL}} \sin\theta$, is mediated solely by the interfacial spin-lattice coupling.
 This is our third main result: the competition between the magnetoelastic coupling and the interfacial spin-lattice coupling leads to the direction-dependent hybridization between magnons and phonons (Fig.~\ref{fig:3}), which in turn leads to nonreciprocal phonon absorption by magnons (Fig.~\ref{fig:4}).

The asymmetric magnon-phonon coupling gives rise to nonreciprocal phonon absorption.
To quantify this effect, we compute the absorption rate of phonon energy transferred to magnons, $\Gamma_\text{m-ph} = {\Delta P}/{\langle {\cal E}_{\text{ph}}\rangle}$,
where $\Delta P = \gamma \rho_s  \langle {\v B}_{\text{m-ph}}(t) \cdot \dot{\v m}(t)\rangle$ and $\langle {\cal E}_{\text{ph}}\rangle$
are the time-averaged absorption power and phonon energy, respectively.
Here, ${\v B}_{\text{m-ph}} = \frac{1}{\gamma \rho_s} \frac{\partial{\cal L}_{\text{m-ph}}}{\partial \v m}$
is the effective magnetic field induced by the magnon-phonon couplings [see Supplemental Material~\cite{SM}].
As shown in Fig.~\ref{fig:4}(a) and (c), the absorption rate is strongly nonreciprocal for $\theta = 0$ but reciprocal at $\theta = \pi/2$.
Figures~\ref{fig:4}(b) and (d) show the corresponding propagation length, $l_{\text{ph}} \sim v_g/(\Gamma_0 + \Gamma_\text{m-ph})$,
where $\Gamma_0 = \omega/Q$ and $Q$ denotes the quality factor characterizing the intrinsic phonon attenuation.
A pronounced directional contrast appears for $\theta=0$, whereas opposite propagation directions become identical at $\theta = \pi/2$.
We emphasize that this nonreciprocal absorption is a generic feature of transverse phonons for $\theta \neq \pi/2$ [see Fig.~\ref{fig:4}(e)].

\emph{Discussion.}\textemdash
We have shown that the Rashba spin-orbit interaction, together with lattice motion, gives rise to a hitherto unrecognized, symmetry-allowed interfacial
spin-lattice coupling.
In the low-frequency limit, we obtained an effective phonon model
in which the interplay between the interfacial spin-lattice coupling and the magnetoelastic coupling generates a kineo-elastic term
that is linear in both the time derivative and the wave vector.
This term produces a finite transverse-phonon velocity asymmetry that can
exceed by orders of magnitude the conventional magnetoelastic scale $\Delta v / v_0 \sim 10^{-5}$.
Beyond the low-frequency limit, the competition between the interfacial
spin-lattice and conventional magnetoelastic couplings produces directional
magnon-phonon hybridization, leading to strong asymmetric phonon absorption
and propagation lengths.

We also comment on the role of the Dzyaloshinskii-Moriya interaction (DMI), which can in general coexist with interfacial Rashba spin-orbit coupling.
As shown in the Supplemental Material~\cite{SM}, DMI does not modify
the leading kineo-elastic term in the adiabatic low-frequency regime,
contributing only higher-order corrections.
In the full magnon-phonon spectrum, by contrast, DMI introduces an additional odd-in-$k$ asymmetry in the magnon dispersion.
Nevertheless, because DMI enters only the magnon sector, the directional magnon-phonon hybridization discussed here remains primarily 
governed by the competition between the interfacial spin-lattice coupling and the conventional magnetoelastic coupling (see Supplemental Material~\cite{SM}).
The same consideration applies to the dipolar interaction, which also acts within the magnon sector; moreover, 
dipolar-induced nonreciprocity is expected to be subdominant in the thin-film geometry considered here.

Our results also suggest an experimentally accessible way to
distinguish the proposed mechanism from previously known sources of
direction-dependent acoustic absorption.
Such absorption asymmetry has been observed in experiments using Rayleigh-waves, 
where conventional magnetoelastic and magneto-rotation interactions can produce a helicity mismatch 
with the magnetization precession~\cite{Xu2020,Kuss2020}.
This contribution can be strongly suppressed by using
shear-horizontal surface acoustic waves (Love waves), whose
out-of-plane motion is strongly reduced, while the present
spin-lattice coupling remains active through the in-plane transverse
lattice motion~\cite{Dreher2012,Yang2024}.
Direction-resolved attenuation measurements using Love waves therefore
provide a natural probe of the directional magnon-phonon
coupling predicted here.

We note that, although the directional magnon-phonon hybridization does not rely on damping, finite Gilbert damping is required for it to appear as a finite absorption rate and a propagation-length contrast. Additional interfacial losses, such as roughness or leakage of the acoustic mode into the substrate, weaken the directional contrast quantitatively but do not alter the mechanism.

Finally, we comment on the range of validity of the microscopic model.
The $p$-orbital construction used here provides a minimal realization
of the coupling; the mechanism itself is not restricted to $p$-orbital systems.
Its essential ingredients are interfacial inversion-symmetry breaking,
orbital hybridization, spin-orbit coupling, and exchange coupling to
the magnetization. Accordingly, the same symmetry-allowed coupling
can arise in other orbital manifolds, including transition-metal
$d$ bands, although its coefficient is material and band dependent
~\cite{Park2013,PKim2014}. We therefore expect this coupling to be relevant to a broad class of magnetic systems exhibiting interfacial Rashba spin-orbit coupling.
For multi-sublattice magnets, analogous local couplings can involve
both the net and staggered magnetic order parameters, suggesting an
interesting extension of the present mechanism to antiferromagnetic
systems.

\section*{Acknowledgement}

We thank Daniel E. Parker for insightful discussions.
G.G. was supported by the National Research Foundation of Korea (NRF-2022R1C1C2006578) and by a Korea University Grant.
S.K.K. was supported by Brain Pool Plus Program through the National Research Foundation of Korea funded by the Ministry of Science and ICT (2020H1D3A2A03099291).

\bibliography{reference}

\end{document}